\documentclass[12pt,a4paper]{article}
\usepackage[latin1]{inputenc}
\usepackage{graphicx,psfrag}
\usepackage{amsmath,amssymb,amsfonts,amsthm}
\usepackage{multirow}
\usepackage{rotating}
\def\n{\nonumber}
\def\be{\begin{equation}}
\def\ee{\end{equation}}
 
\theoremstyle{plain}

\theoremstyle{definition}

\title{Higher dimensional charged shear-free relativistic models with heat flux}
\author{Y Nyonyi\footnote{yusuf@aims.ac.za}, S D Maharaj\footnote{maharaj@ukzn.ac.za} \mbox{} and K S Govinder\footnote{govinder@ukzn.ac.za} \\ Astrophysics and Cosmology Research Unit\\School of Mathematics, Statistics and Computer Science\\University of KwaZulu-Natal\\ Private Bag X54001\\  Durban 4000, South Africa.}
\date{}
\begin{document}
\maketitle

\section*{Abstract}
We analyse shear-free spherically symmetric relativistic models of gravitating fluids with heat flow and electric charge defined on higher dimensional manifolds. The solution to the Einstein-Maxwell system is governed by the pressure isotropy condition which depends on the spacetime dimension. We study this highly nonlinear partial differential equation using Lie's group theoretic approach. The Lie symmetry generators that leave the equation invariant are determined. We provide exact solutions to the gravitational potentials using the first symmetry admitted by the equation. Our new exact solutions contain the earlier results for the four-dimensional case. Using the other Lie generators, we are able to provide solutions to the gravitational potentials or reduce the order of the master equation to a first order nonlinear differential equation. We derive the temperature transport equation in higher dimensions and find expressions for the causal and Eckart temperatures showing their explicit dependance on the dimension. We analyse a particular solution, obtained via group techniques, to show its physical applicability. \linebreak 

\noindent \textbf{Keywords} Gravitating fluids, Exact solutions and Lie symmetries

\section{Introduction}
In this paper, we study charged shear-free spherically symmetric gravitating fluids defined on higher dimensional manifolds. The idea of higher dimensions stems from the earlier attempts of Kaluza $\cite{kk1}$ and Klein $\cite{kk2}$ who were motivated by the desire to unify the fundamental forces of electromagnetism and Einstein gravity by introducing a compact fifth dimension. The discourse of higher dimensional models hibernated for over four decades and it was not until the early 1960's that the early developments to what we have come to know as String Theory came into existence. This theory which requires a higher dimensional framework was initially sought to explain the strong nuclear force but its peculiar properties made it a good candidate for studying quantum gravity with a hope of obtaining a unifying grand theory. In addition, studying models in higher dimensions provides a platform to understand the nature of the early universe. It is believed that the universe, in its earlier epoch was dense and hot (a scenario better explained in higher dimensions), and as a result of expansion the extra dimensions have compactified to produce the present four dimensional universe $\cite{cd1}$.

The model we study here is for a charged higher dimensional shear-free gravitating fluid in the presence of heat flow; this is intended to extend our earlier study $\cite{nmg1}$ in four dimensions. This model is very important in studying both relativistic cosmological and astrophysical processes. Therefore providing exact solutions to the Einstein-Maxwell system is a vital aspect in this regard. This is well documented in Krasinski's  monograph $\cite{ka}$ where he points out the significance of these solutions in understanding the growth of inhomogeneities, the appearance of singularities, structure formation, gravitational collapse and other relativistic stellar processes. Incorporating heat flow and charge in our model provides us with a platform for building radiating and gravitating models. The intricacies of the model we study are simplified by considering the shear-free condition. In this way the generalized pressure isotropy condition reduces to an equation, containing two independent metric functions, which is much easier to study and solve. For a comprehensive recent treatment of shear-free heat conducting perfect fluids see Ivanov  $\cite{bvi}$.

The study of relativistic stars that emit null radiation in the form of radial heat flow, as established by Santos $\cite{sok}$, requires a nonzero heat flux emanating from the interior spacetime to match with the pressure at the boundary with the exterior Vaidya spacetime. This was extended by Maharaj $\textit{et al}$ $\cite{mgg1}$ to the generalized Vaidya spacetime superposing a null fluid and a string fluid in the exterior energy momentum tensor. The Santos junction condition is also applicable to relativistic models in higher dimensions $\cite{bdc1}$. Several models in the higher dimensional setting have been studied over the years with emphasis on understanding gravitational collapse and appearance of naked singularities  $\cite{bdc1,gd1,gb1,jdm1}$. Much of this study is summarised in a systematic manner by Goswami and Joshi $\cite{gj1}$ in their study of higher dimensional spherically symmetric dust collapse; they showed that both black holes and naked singularities would develop as end states depending on the initial data from which the collapse emanates. These studies expound the conditions under which naked singularities may occur in gravitational collapse. However, most of the papers mentioned above are numerically inclined. The existence of an exact analytic model provides a channel to test the accuracy and reliability of numerical solutions. Several attempts have been made over the years to obtain exact solutions in higher dimensions. Bhui $\textit{et al}$ $\cite{bcb1}$ derived the defining Einstein field equations in higher dimensions and used them to study non-adiabatic gravitational collapse. Banerjee and Chatterjee $\cite{bc1}$ provided conditions under which a spherical heat conducting fluid in higher dimensions collapses without the appearance of the horizon. We have previously used the Deng approach to provide new classes of solutions with heat flow while generalizing the four dimensional results  $\cite{nmg2}$. A more systematic approach using group theoretic techniques was adopted by Msomi $\textit{et al}$ $\cite{mgm2}$ to study the same model. They used the Lie analysis of differential equations to generate explicit solutions to the defining pressure isotropy condition. Other studies in this context include the treatment of Ray $\textit{et al}$ $\cite{rbm1}$ who established the existence of an electromagnetic mass distribution corresponding to charged dust in higher dimensions. Hackmann $\textit{et al}$ $\cite{hkkl}$ provided a comprehensive catalogue of analytical solutions of the geodesic equation of massive test particles in higher dimensions in a variety of well known spacetimes. It is clear that models with higher dimensions have a number of important physical applications.

In this study, we consider the general framework of a shear-free higher dimensional charged, heat conducting fluid without placing any restrictions on the spacetime dimension. By applying a group theoretic approach via Lie symmetries, we study the dynamics of the charged shear-free heat conducting model in higher dimensions. We present the Einstein-Maxwell field equations and the generalized pressure isotropy condition in $\S \ref{nmodel}$. A detailed description of the procedure for obtaining the symmetry generators follows in $\S \ref{nlie-analy}$. The first symmetry obtained is used to provide new solutions for any given form of charge in $\S \ref{nusingnx1}$. The gravitational potentials can be found explicitly. In $\S \ref{4sum}$, we summarise the results obtained by using the rest of the symmetries in tabular form. The cases where reduction to quadrature is difficult to perform arises from the nonlinearity of the resultant equations. The results of this paper generalize earlier studies from the four-dimensional manifold to higher dimensions. We consider the temperature and heat transport in higher dimensions, and we generate explicit forms of the temperature in the Eckart theory and the causal theory in $\S \ref{heatTrans}$. Some concluding remarks follow in $\S \ref{discn4}$.

\section{The model} \label{nmodel}
We consider a shear-free spherically symmetric gravitating fluid, in the presence of an electromagnetic field, defined on an $(n+2)-$dimensional manifold. Then the line element in the extended Schwarzschild coordinates $(t,r,\theta_{1}, \dotsc , \theta_{n})$ becomes
\be \label{nline}  \mathrm{d}s^2 = -D^{2}\mathrm{d}t^2 + \frac{1}{V^{2}}\left(\mathrm{d}r^{2} + r^{2}\mathrm{d}X^{2}_{n} \right), \ee
where $n\geq 2$. The gravitational potential components $D$ and $V$ are functions of $r$ and $t$ and 
\be \mathrm{d}X^{2}_{n} = \mathrm{d}\theta^{2}_{1} + \sin^{2}\theta_{1}\mathrm{d}\theta^{2}_{2} +\dots +\sin^{2}\theta_{1}\sin^{2}\theta_{2}\dots \sin^{2}\theta_{n-1}\mathrm{d}\theta^{2}_{n}. \ee

For a charged interior matter distribution, the energy momentum tensor is of the form 
\be \label{ntab} T_{ab} = (\rho + p) U_{a} U_{b} + pg_{ab} + q_{a} U_{b} + q_{b} U_{a}  + E_{ab}, \ee
where $\rho$, $p$, $q_{a} = \left(0,q,0,\cdots,0 \right)$  and $E_{ab}$ are the energy density, the isotropic pressure, the $(n+2)$ heat flux vector and the electromagnetic contribution to the matter distribution respectively. The quantities above are measured relative to a unit, timelike comoving velocity vector $U^{a} = \left(\frac{1}{D},0,\cdots,0 \right)$. 

The nontrivial Einstein-Maxwell equations for the charged gravitating relativistic fluid in comoving coordinates, emanating from equations $\eqref{nline}$ and $\eqref{ntab}$, are
\begin{subequations} \label{nEMcomps}
\begin{align}
\rho &= \frac{n(n+1)V^{2}_{t}}{2D^{2}V^{2}} - \frac{n(n+1)VV^{2}_{r}}{2} + nVV_{rr} +\frac{n^{2}VV_{r}}{r} - \frac{V^{2}}{2D^{2}} \phi^{2}_{r}, \label{nEMcomp1} \\ \n \\
p &= -\dfrac{nD_{r}VV_{r}}{D} + \frac{nD_{r}V^{2}}{rD}+ \frac{n(n-1)V_{r}^{2}}{2} - \frac{n(n-1)VV_{r}}{r} \n \\  
 & \quad + \frac{nV_{tt}}{D^{2}V} - \frac{n(n+3)V_{t}^{2}}{2D^{2}V^{2}} - \frac{nD_{t}V_{t}}{D^{3}V} + \frac{V^{2}}{2D^{2}} \phi^{2}_{r}, \label{nEMcomp2} \\ \n \\
p &= \dfrac{D_{rr}V^{2}}{D} - (n-1)VV_{rr} + \frac{n(n-1)V_{r}^{2}}{2}  + \frac{(n-1)D_{r}V^{2}}{rD} -  \frac{(n-1)^{2}VV_{r}}{r} \n \\ 
&\quad - \frac{(n-2)D_{r}VV_{r}}{D}  + \frac{nV_{tt}}{D^{2}V} - \frac{n(n+3)V_{t}^{2}}{2D^{2}V^{2}} - \frac{nD_{t}V_{t}}{D^{3}V} - \frac{V^{2}}{2D^{2}} \phi^{2}_{r},  \label{nEMcomp3} \\ \n \\
q &= -\frac{nVV_{tr}}{D} + \frac{nV_{r}V_{t}}{D} +\frac{nD_{r}VV_{t}}{D^{2}}, \label{nEMcomp4}   \\ \n \\
0 &= -\dfrac{V^{2}}{D^{2}} \left( \phi_{rt} - \left( (n-1)\frac{V_{t}}{V} + \frac{D_{t}}{D} \right) \phi_{r} \right),  \label{nEMcomp5} \\ \n \\
\sigma &= \dfrac{V^{2}}{D} \left( \phi_{rr} +  \phi_{r} \left( \frac{n}{r} - \frac{D_{r}}{D}  - (n-1)\frac{V_{r}}{V} \right) \right), \label{nEMcomp6}
\end{align} 
\end{subequations} 
where $\sigma$ is the proper charge density. It is important to note that the system $\eqref{nEMcomps}$ contains the results of Nyonyi $\textit{et al}$ $\cite{nmg1}$ when $n=2$.

Integrating $\eqref{nEMcomp5}$ gives $\phi_{r}$ in the form
\be \label{phi-r} \phi_{r} = V^{(n-1)}DF(r), \ee
where $F(r)$ is an arbitrary function. By equating $\eqref{nEMcomp2}$ with $\eqref{nEMcomp3}$ and taking $\eqref{phi-r}$ into consideration, we obtain the higher dimensional generalized pressure isotropy condition 
\be \label{nIsotropy1} 
\dfrac{D_{rr}V^{2}}{D} - (n-1)VV_{rr} - \frac{D_{r}V^{2}}{rD} + 2\frac{D_{r}VV_{r}}{D}-\frac{(n-1)VV_{r}}{r} - \frac{V^{(n-1)}}{D}F(r) = 0. \ee 
The transformation $u =r^{2}$ yields the generalized pressure isotropy condition in $\eqref{nIsotropy1}$ to the simpler form
\be \label{nisotropy} 4u VD_{uu} +8uD_{u}V_{u} - 4u(n-1)DV_{uu} -V^{n} F(u) = 0, \ee 
where the function $F$ now depends on $u$. Knowledge of $F(u)$, $V$ and $D$ solves the Einstein-Maxwell system $\eqref{nEMcomps}$. Therefore we seek to obtain solutions to  equation $\eqref{nisotropy}$ using the Lie analysis, a method that has been previously employed effectively in studying equations in general relativity $\cite{nmg2,mgm2,mgm1}$. It is worth noting that $\eqref{nisotropy}$ reduces to the pressure isotropy condition of a four dimensional charged model with heat flow when $n = 2$:
\be 4u VD_{uu} +8uD_{u}V_{u} - 4uDV_{uu} -V^{2} F(u) = 0, \ee
which was studied by Nyonyi $\textit{et al}$ $\cite{nmg1}$. Equation $\eqref{nisotropy}$ becomes
 \be  VD_{uu} +2D_{u}V_{u} - (n-1)DV_{uu} =0, \ee
in the absence of charge in higher dimensions. This case was comprehensively studied by Nyonyi $\textit{et al}$ $\cite{nmg2}$ and Msomi $\textit{et al}$ $\cite{mgm2}$.
 
\section{Analysis of the problem} \label{nlie-analy}
The nature of the master equation $\eqref{nisotropy}$ enables us to treat it as a second order nonlinear ordinary differential equation in $u$ even though both the potential functions $D$ and $V$ are functions of $u$ and $t$. After integration we introduce the temporal component by taking the constants of integration to be functions of $t$. 

We seek to obtain an infinitesimal generator of the form
\be \label{nInf} X = \xi \partial_{u} + \eta^1  \partial_{D} + \eta^2 \partial_{V}. \ee
that leaves equation $\eqref{nisotropy}$ (hereinafter labelled as $E=0$ for simplicity) invariant. In order to do this we require
\be X^{[2]}E|_{E=0} = 0, \ee
where $X^{[2]}$ is the second prolongation of $X$ required to transform the derivatives in $\eqref{nisotropy}$. 
(See Bluman and Kumei $\cite{bk1}$ and Olver $\cite{o1,o2}$ for details of the standard procedure that is followed here.)

This gives the infinitesimals
\begin{align}
\xi &= C^{0}(u), \label{nxi222}\\ 
\eta^{1} &= c_{1}D, \label{neta111} \\
\eta^{2} &= \left( \frac{c_{1}}{n-1} + c_{2}+ \frac{1}{2}C^{0}_{u} \right) V, \label{neta222}
\end{align}
with a condition on $F(u)$ given by
\be \label{nsys2aa} 2u(n-1)VDC^{0}_{uuu} + V^{n} \left[ F \left( -\frac{C^{0}}{u} + \frac{(n-1)+4}{2}C^{0}_{u} + (n-1)c_{2}\right)+C^{0}F^{\prime} \right] = 0. \ee
When $F$ is arbitrary, equation $\eqref{nsys2aa}$ is satisfied if both
\be \label{nco-arbit-f} C^{0}(u) = 0, \ee
and 
\be \label{nc2-arbit-f} c_{2} = 0. \ee
Using $\eqref{nxi222}$--$\eqref{neta222}$ and $\eqref{nco-arbit-f}$--$\eqref{nc2-arbit-f}$ the infinitesimals become
\begin{align}
\xi(u) &= 0, \\ 
\eta^{1}(D) &= c_{1} D, \\
\eta^{2}(V) &= \frac{c_{1}}{n-1} V .
\end{align}
Following $\eqref{nInf}$, we observe that the infinitesimal generator 
\be X_{1} = D \partial_{D} + \frac{V}{n-1} \partial_{V}, \label{nx1} \ee
is the only symmetry admitted by $\eqref{nisotropy}$ when $F(u)$ is arbitrary. It is remarkable that $\eqref{nisotropy}$ admits a symmetry in general without placing any restriction on the form of the charge. This symmetry depends on the dimension $n$, and reduces to the four dimensional case when $n=2$. 

By taking $\eqref{nsys2aa}$ to be a restriction on $F$ we obtain two conditions
\be \label{nC00} C^{0}_{uuu} = 0, \ee 
and 
\be \label{nfu} F \left( -\frac{C^{0}}{u} + \frac{n+3}{2}C^{0}_{u} + (n-1)c_{2}\right) + C^{0}F^{\prime} = 0, \ee
Solving equation $\eqref{nC00}$ gives 
\be \label{nC001} C^{0}(u) = c_{3}u^{2} + c_{4}u + c_{5}. \ee
We then solve $\eqref{nfu}$ using $\eqref{nC001}$ to obtain
\be  \label{nfgeneral} F(u) = \dfrac{uc_{6}}{(c_{3}u^{2} + c_{4}u + c_{5})^{(n+3)/2}}  \exp \left[ \dfrac{2(n-1)c_{2}}{\sqrt{-c_{4} + 4c_{3} c_{5}}} \arctan \left(\dfrac{c_{4}+2c_{3}u}{\sqrt{-c_{4} + 4c_{3}c_{5}}} \right) \right],  \ee
 as the definition of $F(u)$. Note that $c_{6}$ is a constant of integration. From $\eqref{nC001}$--$\eqref{nfgeneral}$, we observe that $\eqref{nisotropy}$ admits another symmetry
\be \label{nx2} X_{2} =  \left(c_{3}u^{2} + c_{4}u + c_{5} \right) \partial_{u} + \left(c_{2} + c_{3}u + \frac{c_{4}}{2}\right)V \partial_{V}, \ee
when $F(u)$ takes on the form $\eqref{nfgeneral}$.

Equation $\eqref{nfgeneral}$ is the most general higher dimensional form of $F(u)$ for which $X_{2}$ is the corresponding symmetry. Simpler forms of $F(u)$ can be obtained from the constants existing in $\eqref{nfgeneral}$ that relate to the symmetry $X_{2}$. In addition the simpler higher dimensional forms of $F(u)$ may admit extra symmetries. This is summarised in Table $\ref{4nfs}$. When $c_2 = 0$ (with all other constants nonzero), we obtain an extra symmetry as indicated in the last row of Table $\ref{4nfs}$. Msomi $\textit{et al}$ $\cite{mgm2}$ comprehensively studied the higher dimensional uncharged case ($F=0$). It is important to further highlight that verification of the generator $X_6$ as a symmetry to our master equation for the uncharged case revealed that it is independent of the dimension $n$. This serves to correct the result published by Msomi $\textit{et al}$ $\cite{mgm2}$.
\begin{sidewaystable}[h!]
\caption{Symmetries associated with different forms of $F(u)$}
\label{4nfs}
\begin{center}
\begin{tabular}{l l l}
\hline Symmetry generator & Form of $F(u)$ & Extra symmetries \\ 
\hline \multirow {4}*{$X_{2,1} = \frac{V}{n-1} \partial_{V}$}  &\multirow {4}*{$0$} & $X_{3} =\partial_{u}$,  \\
    &      &$X_{4} = u\partial_{u}$,\\
    &      &$X_{5}=D\partial_{D} $, \\
    &      &$X_{6} = u^{2} \partial_{u} + uV \partial_{V}$  \\ 
\hline  $X_{2,2} = u^2\partial_{u} +uV\partial_{V}$  & $k u^{-(n+2)}$ & $X_{3}= (n-1)u\partial_{u} +(n+1)V\partial_{V}$ \\ 
\hline  $X_{2,3} = u\partial_{u} +\frac{1}{2} V\partial_{V}$  & $ ku^{-\frac{n+1}{2}}$ & None \\ 
\hline  $X_{2,4} = \partial_{u}$  & $ k u$ & $X_{3} = -(n-1)u\partial_{u} +2V\partial_{V}$\\
\hline $X_{2,5} = \left(c_{3}u^{2} + c_{4}u + c_{5} \right)\partial_{u}+ \left(c_{3}u + \frac{c_{4}}{2} \right)\partial_{V}$ &  $ c_{6}(c_{3}u^{2} + c_{4}u + c_{5})^{-(n+3)/2}$ & None \\ 
\hline 
\end{tabular}
\end{center} 
\end{sidewaystable}

\section{New solutions using symmetries}
It is well documented that symmetries of differential equations are used to reduce the order of the equation with the hope of obtaining simplified forms which then can be solved. In this section, we intend to give a detailed description of the reduction of our master equation $\eqref{nisotropy}$ using $X_{1}$. For the remaining symmetries, we will summarise the results in tabular form.
\subsection{Arbitrary $F$} \label{nusingnx1}
Since $F(u)$ is arbitrary, our master equation to be analyzed is simply 
\be \label{narbtf} VD_{uu} +2V_{u}D_{u} -(n-1) DV_{uu} - V^{n} \frac{F(u)}{4u} = 0. \ee
It is obvious that, to make any headway, one has to assume a relationship between the metric functions $D$ and $V$.  However, such an {\it ad hoc} approach is sure to cause frustration for all but the most gifted of practitioners. We are fortunate that we can make recourse to the fact that this equation does admit a Lie point symmetry regardless of the form of $F(u)$.

We proceed by determining that
the associated Lagrange's system for the first extension of
\be X_{1} = D\partial_{D} + \frac{V}{n-1} \partial_{V} \ee
is given by
\be \dfrac{\mathrm{d}u}{0} = \dfrac{\mathrm{d}D}{D} = \dfrac{\mathrm{d}V}{V/(n-1)} = \dfrac{\mathrm{d}D^{\prime}}{D^{\prime}} = \dfrac{\mathrm{d}V^{\prime}}{V^{\prime}/(n-1)}. \ee
The invariants are then found to be 
\begin{subequations}
\begin{align}
 p &= u, \\
q(p) &= \dfrac{V}{D^{1/(n-1)}}, \\ 
r(p) &= \dfrac{D^{\prime}}{D}, \\
s(p) &= \dfrac{V^{\prime}}{D^{1/(n-1)}}.
\end{align}
\end{subequations}
While this is the full set of first order invariants obtained, we only use $p$, $q$ and $r$. Substituting these invariants into $\eqref{narbtf}$ yields
\be q^{\prime \prime} = - \dfrac{F(p)}{(n-1)4p} q^{n}  + \frac{n}{(n-1)^2} qr^{2}.\label{nrpqua}  \ee
If we had not used the symmetry $X_1$, there would have been little, if any, hope that the variables chosen would have led to an equation solely in terms of those variables as was obtained here.  This is one of the great uses of Lie symmetries.

We can treat  equation $\eqref{nrpqua}$ as a definition for $r$ and obtain
\be \label{nrpqu} r = \pm \sqrt{\left[\dfrac{q^{\prime \prime}}{q} +\frac{q^{n-1}}{n-1} \frac{F(p)}{4p} \right]\dfrac{(n-1)^2}{n} }. \ee
Reverting to the original variables produces
\be \dfrac{D^{\prime}}{D} = \pm \sqrt{\dfrac{(n-1)^2}{n} \left[\dfrac{q^{\prime \prime}}{q} +  \frac{q^{n-1}}{n-1} \frac{F(p)}{4p} \right]}.\ee
We can now integrate this equation to obtain
\be \label{nsoln1} D = \exp \left(\pm C \int  \sqrt{\dfrac{(n-1)^2}{n} \left[\dfrac{W^{\prime \prime}}{W} +  \frac{W^{n-1}}{n-1}\frac{F(u)}{4u}\right]} \quad \mathrm{d}u \right), \ee
where $C$ is a constant of integration.

Equation $\eqref{nsoln1}$, illustrates that whenever we are given any generalized ratio of the gravitational potentials $W=\frac{V}{D^{1/(n-1)}}$ and an arbitrary function $F(u)$ indicative of charge, we can explicitly obtain the exact expression for the potentials. We have thus obtained a generating function approach for solving $\eqref{nsoln1}$, something that, to the best of our knowledge has not been obtained before in higher dimensions. When we set $n = 2$, we regain 
\be  D = \exp \left[\pm C \int  \sqrt{\dfrac{W^{\prime \prime}}{2W} +  W \dfrac{F(u)}{8u}} \quad \mathrm{d}u \right]. \ee
This is the result for the four-dimensional model studied by Nyonyi $\textit{et al}$ $\cite{nmg1}$. Also, when we set $F(u)=0$ in $\eqref{nsoln1}$, we obtain 
\be \label{nsolnmmg} D = \exp \left[\pm C \int  \sqrt{\dfrac{(n-1)^2}{n} \dfrac{W^{\prime \prime}}{W}}  \quad \mathrm{d}u \right]. \ee
This is the uncharged solution of Msomi $\textit{et al}$ $\cite{mgm2}$. The case for $n=2$ in $\eqref{nsolnmmg}$ was obtained by Msomi $\textit{et al}$ $\cite{mgm1}$.

\subsection{Summary using other symmetries} \label{4sum}
The solutions associated with the symmetry $X_1$ are contained in $\S \ref{nusingnx1}$. In this section we discuss the solutions obtained by using the symmetries emanating from the specified forms of $F(u)$. The symmetries relating to the simpler forms of $F(u)$ (as described in Table $\ref{4nfs}$) are used to reduce the order of the equation with the hope of obtaining new solutions. Table $\ref{4slnsym}$ summarises the symmetries (obtained with their corresponding form of $F(u)$) that we have used to provide solutions to the respective master equations. In all cases we obtained the invariants from the first prolongation of the symmetries. This is highlighted in the third column of Table $\ref{4slnsym}$. Using a partial set of invariants, we are able to demonstrate the existence of exact solutions to the respective master equations for the specified simpler forms of $F(u)$ (as illustrated in the fourth column of Table $\ref{4slnsym}$). However, not all symmetries we obtained were able to provide exact solutions to the master equation. For these symmetries, we were able to reduce the order of the equation. The reduced forms we obtained were difficult to reduce to quadrature. These are summarised in Table $\ref{4rdcdsym}$.
\begin{sidewaystable}[h!]
\caption{Solutions from other symmetries (excluding $X_1$): $A$, $B$ are constants of integration, $\alpha_n =((n+1)/(n-1))$ }
\label{4slnsym}
\begin{center}
\begin{tabular}{l l l l}
\hline $F(u)$ & Symmetry & Invariants  & Solution \\ 
\hline  \multirow {3}*{ 0 }   &  \multirow {3}*{$\frac{V}{n-1}\partial_{V}$} & $p = u$, $q(p) = D$,& $D = (n-1)$ \\
           &        & $r(p)= D^{\prime}$, &\multirow {2}*{$\mbox{ }\quad\int\left(e^{-2 ( V^{\prime}/V)} \int D \left[ \dfrac{\mathrm{d}(V^{\prime}/V)}{\mathrm{d}u} + \left( \dfrac{ V^{\prime} }{V} \right)^{2} \right]  e^{2(V^{\prime}/V) } \mathrm{d}u \right) \mathrm{d}u $} \\
           &          &  $ s(p) = \dfrac{V^{\prime}}{V}$ &  \\
\hline \multirow {4}*{$k u^{-(n+2)}$} & $u^2 \partial_{u} +uV\partial_{V}$ &  $p=D$,  &  $\int \left(2\frac{2 \frac{\mathrm{d}(V/u)}{\mathrm{d}D} - (n-1)D \frac{{\mathrm{d}}^2(V/u)}{{\mathrm{d}D}^2} }{(V/u)-  (n-1)D \frac{\mathrm{d}(V/u)}{\mathrm{d} D} } \right)^{1/2}$ \\ 
           &          & $q(p)=\frac{V}{u}$, & \multirow {2}*{$\sqrt{[\exp[2\frac{2 \frac{\mathrm{d}(V/u)}{\mathrm{d}D} - (n-1)D \frac{{\mathrm{d}}^2(V/u)}{{\mathrm{d}D}^2} }{(V/u)-  (n-1)D \frac{\mathrm{d}(V/u)}{\mathrm{d} D}}(D + A)] +\frac{(k/4) (V/u)^{2}}{(V/u)-  (n-1)D \frac{\mathrm{d}(V/u)}{\mathrm{d} D}}]^{-1}} \mbox{ } \mathrm{d} D$ } \\
           &          &  $r(p) = u^{2} D^{\prime}$,  &  \\
           &          & $s(p) = u\left(q(p)- V^{\prime} \right)$  & $ = -\frac{1}{u} + B$ \\
\hline \multirow {4}*{$k u^{-(n+2)}$} & \multirow {4}*{$(n-1)u \partial_{u} +(n+1)V\partial_{V}$} & $p=D$, &  \multirow {2}*{$\int -\frac{Y/(2[D-C])}{\exp[Y]+ (VX/((n-1)Du^{\alpha_n}))} = \frac{u^{\alpha_n}}{\alpha_n} + A,$  }\\
           &          & $q(p)=\frac{V^{(1/(n+1))}}{u^{(1/(n-1))}}$, & \\
           &          & $r(p) = uD^{\prime}$,  & where $X=\frac{uD^{\prime \prime }}{D^{\prime}} - \frac{kV^{(n-1)}}{4D^{\prime}u^{(n+2)}} $  \\
           &          & $s(p) = \sqrt{V^{\prime}}u^{(-1/(n-1))}$ & $\qquad \quad Y=\frac{4(uD^{\prime} - D)(D-C)}{(n-1)uD^{\prime}D} $ \\
\hline \multirow {4}*{$ku^{-\frac{(n+1)}{2}} $}  & \multirow {4}*{$ u\partial_{u} +\frac{1}{2} V\partial_{V}$} & $p= D$,  & \multirow {2}*{$\int \sqrt{X} \left(\exp[X(D+A)] -Y\right)^{-1/2} \mathrm{d} D = \ln u + B$ } \\
           &          & $q(p)= \frac{V^{2}}{u}$, &  \\
           &          & $r(p) = u D^{\prime}$,   & where $X=\frac{4\frac{\mathrm{d}(V^2/u)}{\mathrm{d} D}  - (n-1)D \left[ 2\frac{{\mathrm{d}}^{2}(V^2/u)}{{\mathrm{d} D}^2} - \left(\frac{\mathrm{d}(V^2/u)}{\mathrm{d} D}\right)^{2} \right]}{2(V^2/u) - (n-1)D \frac{\mathrm{d}(V^2/u)}{\mathrm{d} D}} $ \\
           &          & $s(p) = u {V^{\prime}}^{2}$ & $\qquad \quad Y=  \frac{(n-1)D(V^2/u)-k(V^2/u)^{((n+1)/2)}}{2(V^2/u) - (n-1)D \frac{\mathrm{d}(V^2/u)}{\mathrm{d} D}}$ \\
\hline \multirow {3}*{$ku $} & \multirow {3}*{$ \partial_{u}$} & $p = D$, & \multirow {2}*{ $D = \pm \int  \sqrt{ X} \left(Y +\exp[X(D+k)] \right)^{-1/2} \mbox{ }\mathrm{d}D = u+B$ }   \\ 
           &          & $q(p) = V$,         &   \\
           &          & $r(p) = D^{\prime}$,& where $X= 2\dfrac{2(\mathrm{d}V/\mathrm{d}D) - (n-1)D (\mathrm{d}^{2} V/\mathrm{d}D^2)}{V - (n-1)D(\mathrm{d}V/\mathrm{d}D)} $   \\
           &          & $s(p) = V^{\prime}$ & $\qquad \quad Y= \dfrac{2AV^{n}}{V -(n-1)D(\mathrm{d}V/\mathrm{d}D)}$ \\
\hline   
\end{tabular}
\end{center} 
\end{sidewaystable}       
           
\begin{sidewaystable}[h!]
\caption{Reductions using symmetries for cases not reducible to quadrature: $\beta(u) =c_{3}u^{2} + c_{4}u + c_{5}$ }
\label{4rdcdsym}
\begin{center}
\begin{tabular}{l l l l}
\hline $F(u)$ & Symmetry & Invariants  & Reduced equation \\
\hline \multirow {5}*{$ku $} & \multirow {5}*{$-(n-1)u\partial_{u}+2V\partial_{V}$} &           $p = D$,  &  \multirow {5}*{$r_{p}  - \dfrac{1}{r} \left(\dfrac{(n-1)p t^{4}}{q^{2}} + \dfrac{kq^{2}}{4} \right) + 2\dfrac{s^{(n+1)}}{q^2} - 1 = 0$}\\
           &           & $q(p) = \sqrt{V} u^{(1/(n-1))}$, &  \\
           &           & $r(p) = u D^{\prime}$, &   \\
           &           & $s(p) = \sqrt[n+1]{V^{\prime}} u^{(1/(n-1))}$, &   \\
           &           & $t(p) = \sqrt[2n]{V^{\prime \prime}} u^{(1/(n-1))}$ &  \\
\hline  \multirow {4}*{$k(\beta(u))^{-(n+3)/2}$} & \multirow {4}*{$\beta(u)\partial_{u} + \left(c_{3}u + \frac{c_{4}}{2} \right)\partial_{V}$}  & $ p = D$,& \multirow {4}*{$r_{p} - \frac{1}{r^{2}} \left[ \dfrac{(n-1)t^{2} p}{q^{2}} + \dfrac{q^{(n+1)}c_{6}}{4} \right] + \left( \frac{2s}{q} -c_{4} \right) = 0$} \\
           &            & $q(p) = \frac{V}{\sqrt{\beta(u)}}$,  &      \\ 
           &           & $r(p) = V \sqrt{D^{\prime}}$, & \\          
           &           & $s(p) = -q(p)uc_{3} + V^{\prime}\sqrt{\beta(u)}$,  &   \\
           &           &  $t(p) = V \sqrt[3]{V^{\prime \prime}}$ &  \\ 
\hline
\end{tabular}
\end{center} 
\end{sidewaystable}

It is apparent that we are able to generate a number of solutions to our pressure isotropy equation $\eqref{nisotropy}$.  However, simply generating solutions without context is not a useful exercise.  As a result, we now focus on the physical applicability of our solutions.

\section{Temperature and heat transport} \label{heatTrans}
We now study the causal heat transport equation of  Maxwell-Cattaneo type without viscous stress and rotation. This is given by
\be \label{hteqn} \tau {h_{a}^{ }}^{b} {\dot{q}}_{a} + q_{a} = - \kappa\left({h_{a}^{ }}^{b} T_{;b} + T \dot{u}_{a} \right), \ee
where $\tau(\geq 0)$ is the relaxation time associated with heat transport, $q_{a}$ is the heat flux, $h_{ab} = g_{ab} + u_{a}u_{b}$ is the projection tensor, $T$ is the temperature, $\kappa(\geq 0)$ is the coefficient of thermal conductivity, and $u_{a}$ is the velocity vector. In order to solve $\eqref{hteqn}$ we require knowledge of $\tau$ and $\kappa$. The coefficient of thermal conductivity $\kappa$ is obtained from the interaction between a radiating fluid and matter $\cite{ws1}$. Following the treatment of Mart\'{i}nez $\cite{mj1}$, we now take 
\be \kappa = \gamma T^{3} \tau_{c}, \ee
where $\gamma$ is a constant and $\tau_{c}$ is the mean collision time. On physical grounds we can assume that 
\be \tau = \left(\frac{\beta \gamma}{\alpha} \right) \tau_{c} = \beta T^{-\sigma}, \ee
with $\alpha$, $\beta$ and $\sigma$ are positive constants. Then for the metric $\eqref{nline}$, the transport equation $\eqref{hteqn}$ becomes
\be 
\beta T^{-\sigma} \left(\frac{q}{V} \right)_{,t} + D\left(\frac{q}{V} \right) =  \alpha V T^{3-\sigma} (DT)_{,r}. \ee
It would seem that the higher dimensional nature of our metric does not manifest itself in this equation.  However, when we substitute for the heat flux, our equation becomes
\be \label{nhteqn} n\beta T^{-\sigma} \left(V\left(\frac{V_{t}}{DV}\right)_{,r} \right)_{,t} + nDV\left(\frac{V_{t}}{DV}\right)_{,r} =  \alpha V T^{3-\sigma} (DT)_{,r} \ee
in which the higher dimensional dependence is explicit.  To our knowledge, this is the first statement of a  higher dimensional spherically symmetric causal transport equation.
 We observe that the temperature of the fluid is directly affected by the dimension $n$. Note that $\eqref{nhteqn}$ reduces to the four dimensional case studied by Govinder and Govender $\cite{gg2}$ when $n=2$. 
 
 It is possible to integrate $\eqref{nhteqn}$ under particular assumptions on $\beta$ and $\sigma$. 
The case $\beta = 0$ corresponds to the noncausal Eckart theory. The explicit noncausal expressions for the temperature are
\begin{align}
\ln(DT) &= -\frac{1}{\alpha} \int \left(\frac{q}{V^2}\right) \mathrm{d}r  + G(t), \qquad \sigma =4 \nonumber \\
        &= \frac{n}{\alpha} \left(\frac{V_{t}}{DV}\right) + G(t),\label{nhteqn1} \\ \nonumber \\
(DT)^{4-\sigma} &= \frac{\sigma -4}{\alpha} \int D^{4-\sigma} \left( \frac{q}{V^2}\right) \mathrm{d}r + G(t), \qquad \sigma \neq 4 \nonumber\\
                &= \frac{n(4-\sigma)}{\alpha} \int D^{4-\sigma} \left(\frac{V_{t}}{DV}\right)_{,r} \mathrm{d}r + G(t), \label{nhteqn2}
\end{align}
where $G(t)$ is an arbitrary constant of integration.

The mean collision time is constant when $\sigma = 0$. In this case equation $\eqref{nhteqn}$ simplifies substantially. We obtain
\begin{align}
 (DT)^{4} &= -\frac{4}{\alpha}\left[ \beta \int \frac{D^3}{V} \left( \frac{q}{V}\right)_{,r} \mathrm{d}r + \int D^{4} \left( \frac{q}{V}\right) \mathrm{d}r \right] + G(t) \nonumber \\
          &= \frac{4n}{\alpha}\left[ \beta \int \frac{D^3}{V} \left(V\left(\frac{V_{t}}{DV}\right)_{,r} \right)_{,t} \mathrm{d}r + \int D^{4} \left(\frac{V_{t}}{DV}\right)_{,r} \mathrm{d}r \right] + G(t). \label{nhteqn3} \end{align}
The other case for which we can find the causal temperature explicitly is $\sigma = 4$. The transport equation $\eqref{nhteqn}$ can be solved to give
\begin{align}
(DT)^{4} &= \exp\left( -\frac{4q}{\alpha V} \right) \left[ -\frac{4\beta}{\alpha} \int D^{3} \left(\frac{q}{V} \right)_{,t} \exp\left(\frac{4q}{\alpha V} \right) \mathrm{d}r + G(t) \right] \nonumber \\
         &= \exp\left( \frac{4n}{\alpha}V \left(\frac{V_t}{DV}\right)_{,r} \right) \n \\
         & \quad \times \left[ \frac{4n\beta}{\alpha} \int D^{3} \left(V \left(\frac{V_t}{DV}\right)_{,r}\right)_{,t} \exp\left( -\frac{4n}{\alpha}V \left(\frac{V_t}{DV}\right)_{,r} \right)  \mathrm{d}r + G(t) \right] \label{nhteqn4}
\end{align}
We point out that $\eqref{nhteqn1}$--$\eqref{nhteqn4}$ for the heat transport equations ($n \geq 2$) generalize the results obtained by Govinder and Govender $\cite{gg2}$ when $n=2$.

To study the temperature profiles, it is necessary to complete the integration in $\eqref{nhteqn1}$--$\eqref{nhteqn4}$ and express $T$ in terms of simple functions. We seek to illustrate this property by choosing a suitable example. By considering the class of solutions represented by $\eqref{nsoln1}$, we consider the case when the generalized ratio of gravitational potential $W=u$, the constant of integration $C=1$ and the arbitrary form of charge $F(u)=u^{-n}$. We obtain the gravitational potentials in the form
\be \label{eg} D(t,r) = A r^{2k}, \qquad V(t,r) = r^{2} \left( A r^{2k} \right)^{1/(n-1)}, \ee
where $u=r^2$, $k=\pm \frac{1}{2} \sqrt{\frac{n-1}{n}}$, and $A$ is an arbitrary function of $t$. Note that $V$ is obtained from the expression for the generalized ratio $W=V/D^{(1/(n-1))}$ for the gravitational potentials. 

Following $\eqref{nhteqn1}$ the noncausal exact solution is
\be T = \dfrac{1}{A r^{2k}} \exp \left[\frac{n}{\alpha(n-1)} \frac{A_{t}}{A^{2} r^{2k}} + G(t) \right], \ee
for the case $\sigma = 4$,  and when $\sigma \neq 4$, we obtain the noncausal profile
\be \left(A r^{2k} T \right)^{\sigma-4} = \dfrac{-kn(4-\sigma)}{\alpha(n-1)(3k - k\sigma +1)} A_{t} A^{2-\sigma} r^{(3k - k\sigma + 1)} + G(t), \ee
using $\eqref{nhteqn2}$. Further, we can also obtain the causal solutions of the temperature profile. We find that
\be T^{4} = \dfrac{4n}{\alpha(n-1)A^2} \left[ \frac{\beta}{2} A^{((n-2)/(n-1))} \left(A_{t} A^{((1-2n)/(n-1))}\right)_{t} r^{-4k} - \frac{A_{t}}{3} r^{-2k} \right] + \dfrac{G(t)}{A^{4}r^{8k}}, \ee
using $\eqref{nhteqn3}$ for $\sigma=0$. The form for the causal temperature resulting from $\eqref{nhteqn4}$ when $\sigma=4$ is more complicated and we omit this expression. Other choices of the generalized ratio of the gravitational potentials $W$ may lead to forms that yield a more tractable form for $T$ when $\sigma=4$. Our example, for the potentials $\eqref{eg}$, shows that both causal and noncausal temperatures may be found explicitly for the class of models presented in this paper with higher dimensions. It is interesting to note that the temperature profiles can be found exactly in terms of elementary functions and are dependant on the dimension $n$.

\section{Conclusion}\label{discn4}
We have obtained new exact solutions to the generalized Einstein-Maxwell system of charged relativistic fluids in the presence of heat flux defined on higher dimensional manifolds. Our focus was on the generalized pressure isotropy condition. We were able to transform the master equation into a second order nonlinear differential equation that generalized the four-dimensional case. This equation was analysed via a group theoretic approach. 

Interestingly, we were able to find a Lie symmetry 
\be \nonumber X_{1} = D\partial_{D} + \frac{V}{n-1} \partial_{V} \ee
without restricting the electromagnetic field; the form for the charge is arbitrary. The attempt to provide solutions to the generalised Einstein-Maxwell system using this symmetry yielded a remarkable result where exact expressions for the potentials $D$ and $V$ were obtained explicitly when the ratio of the gravitational potentials $W$ coupled with the dimension of the manifold $n$ is known. This general result is an extension of the Nyonyi $\textit{et al}$ $\cite{nmg1}$ model for the four-dimensional case and reduces to the Msomi $\textit{et al}$ $\cite{mgm2}$ solution in the absence of charge. This result would not have been possible without utilising the Lie symmetry approach.

When the electromagnetic field was restricted and the charge takes on a specific form, a second Lie generator 
\be \nonumber X_{2} =  \left(c_{3}u^{2} + c_{4}u + c_{5} \right) \partial_{u} + \left(c_{2} + c_{3}u + \frac{c_{4}}{2}\right)V \partial_{V}, \ee
arose. The functional dependence of the charge distribution for the symmetry $X_2$ is given by $\eqref{nfgeneral}$. In addition, other specific forms of the charge distribution yielded additional symmetries; these are identified in Table $\ref{4nfs}$. We note that our results corrected one of the symmetries presented by Msomi $\textit{et al}$ $\cite{mgm2}$; the symmetry in question is independent of dimension. Note that, using these symmetries, we were able to provide explicit expressions for the gravitational potentials as summarised in Table $\ref{4slnsym}$. For the cases where reduction to quadrature was difficult, we were able to reduce the order of the master equation (see Table $\ref{4rdcdsym}$). The solutions presented in this paper are new and have not been published before. 

Importantly, we also obtained the generalised heat transport equation for the causal and noncausal (when $\sigma = 0$) temperature  
\be \nonumber n\beta T^{-\sigma} \left(V\left(\frac{V_{t}}{DV}\right)_{,r} \right)_{,t} + nDV\left(\frac{V_{t}}{DV}\right)_{,r} =  \alpha V T^{3-\sigma} (DT)_{,r}, \ee
which depends on the dimension of the manifold. This expression reduces to the four-dimensional Govinder and Govender $\cite{gg2}$ result. We demonstrated existence of explicit forms of the temperature $T$ using the gravitational potentials derived from the Lie symmetry analysis. These solutions can be applied to both the uncharged and charged matter distributions defined on a higher dimensional manifold. All these solutions were obtained as a direct result of the symmetry approach.

Further work in this regard includes relaxing the shear-free requirement on the space-time. The analysis of the resulting, complicated, equations is currently underway.

\section{Acknowledgement}
YN, KSG and SDM thank the National Research Foundation and the University of KwaZulu-Natal for financial support. SDM further acknowledges that this research is supported by the South African Research Chair Initiative of the Department of Science and Technology

\end{document}